# Effect of ballistic re-solution on the nucleation kinetics of precipitates in diluted binary alloys under irradiation. Part 1: Stoichiometric γ′ precipitates in Ni-Al alloys

M.S. Veshchunov[*)]

Nuclear Safety Institute (IBRAE), Russian Academy of Sciences,

52, B. Tulskaya, Moscow 115191, Russian Federation

**Abstract**

A modification of classical nucleation theory is carried out as applied to solid solutions under irradiation, taking into account the influence of ballistic re-solution on the nucleation kinetics of pure unary (single-component) and stoichiometric binary precipitates. The effects of excessive point defects formed under steady irradiation conditions and operating alongside the ballistic re-solution mechanism are incorporated into the new model for a consistent description of the nucleation of incoherent and coherent particles. The developed model was applied to interpret the results of Nelson, Hudson and Mazey (NHM) tests, in which the stability of γ′ phase ($Ni_3Al$) precipitates in diluted Ni-Al alloys under irradiation was studied.

[*)] Corresponding author. E-mail: msvesh@gmail.com

## 1. Introduction

It is now experimentally established that irradiation of alloys by energetic particles may not only accelerate solid state phase transformations, but also restrain the formation of new phase precipitates [1]. For instance, Cawthorne and Brown [2] have reported their findings from an electron metallographic examination of stainless steel cladding that has been irradiated in DFR reactor. The investigation revealed that following irradiation at temperatures reaching 700°C, the sigma phase was not identified, although the presence of this phase at these temperatures is consistent with the equilibrium phase diagram and is typically observed during the annealing process.

A similar phenomenon has been observed in W-25Re binary alloys exposed to neutron radiation in EBR-II reactor in the range of $6.1$ to $37 \cdot 10^{21}$ n/cm$^2$ ($> 0.1$ MeV) by Williams, Stiegler and Wiffen [3]. It was reported that electron diffraction patterns exhibited the presence of the cubic χ phase (WRe$_3$) following irradiation, despite the anticipated precipitation of the sigma phase in the alloy. This effect has also been reported by Sikka and Moteff [4] in We-25Re alloy specimens irradiated to fluence of $\sim 1 \cdot 10^{22}$ n/cm$^2$ in EBR-II.

Holloway and Stubbins [5] applied equilibrium thermodynamics to various binary systems with an excess of vacancies to calculate the phase boundary shifts of the equilibrium phase diagram as a function of defect concentration and showed that point defects can strongly affect the solubility limit of a solid solution. For instance in the system Fe-Cr, they found that the range of stability for the sigma phase, which is normally stable between 773 and 1093 K, was considerably reduced owing to the increased concentration of point defects in solid solution.

Schuler et al. [6], using Low Temperature Expansion (LTE) technique to compute equilibrium properties of alloys, demonstrated for three binary systems, Fe-C, Fe-N, and Fe-O, that under irradiation, interstitial solutes can be stabilized in the solid solution because they form vacancy-solute clusters owing to their attractive binding energy with vacancies. It was concluded that the effect of a steady state non-equilibrium vacancy concentration is to increase the solubility limit by altering the solute formation energy in the solid solution. In addition, they incorporated into the analysis an additional mechanism proposed by Martin [7], which involves the enhancement of solute solubility through ballistic mixing of atoms in solid solution and may operate in parallel with the first mechanism. The non-equilibrium solubility is then higher than the equilibrium solubility, so partial dissolution of equilibrium precipitates was anticipated.

The nucleation of new phase precipitates in a solid solution oversaturated with vacancies was modelled by Russell [8], who showed that for an incoherent single-component precipitate with a greater atomic volume $v$ than the matrix $v_m$ (characterised by the volume mismatch parameter $\varphi = (v - v_m)/v_m > 0$) vacancies may act as the second chemical component in nucleation reactions and thereby relieve transformation strains. As a consequence, he concluded that excess vacancies destabilise undersized precipitate phases and stabilise oversized precipitates. The nucleation kinetics in such systems was reconsidered within the framework of Reiss' theory of binary nucleation [11] in the author's work [9, 10], which confirmed Russel's conclusion on the shift of the critical supersaturation, but corrected the nucleation rate, which in [8] was underestimated by several orders of magnitude due to the application of the unary (single-component) nucleation theory to the binary system.

However, for irradiation conditions, Maydet and Russel [12] showed that the effect of vacancies is strongly attenuated by interstitials generated together with vacancies, and therefore often becomes insufficient to explain experimental observations (see also Russel's review paper [13]). In particular, as noted by Kaufman et al. [14], the atomic volume of the sigma phase is higher than that of the stainless steels matrix (studied in [2]), and therefore, according to the Maydet-Russell mechanism, irradiation favours rather than attenuates it, unless another phase with an even larger volume is formed instead. For this reason, it was assumed in [14] that irradiation inhibits the formation of some phases by bombardment processes which physically knock precipitate atoms into solution. To confirm this assumption, the influence of such a mechanism on the nucleation kinetics of precipitates and on the range of stability of the precipitate phase under irradiation will be studied in the present work.

The nucleation of pure unary (single-component) and stoichiometric binary phases will be considered in Part 1 (i.e. in this article), where the new model will also be applied to analyse the behaviour of the stoichiometric $\gamma'$ ($Ni_3Al$) phase precipitates in irradiated Ni-Al alloys studied in experiments of Nelson, Hudson and Mazey [15]. The extension of the theory to the nucleation of non-stoichiometric binary phases will be presented in Part 2 (next article), where the behaviour of Cr-rich $\alpha'$ phase precipitates in irradiated Fe-Cr alloys will be analysed.

## 2. Ballistic re-solution from precipitates

The effect of the ballistically driven re-solution of solute atoms from $Ni_3Al$ precipitates into the surrounding matrix was studied in Ni-Al alloys by Nelson, Hudson and Mazey (NHM) [15], who developed a model for dissolution of precipitates by recoil and disordering processes (cf. also [16]). In this model, for a damage rate in the solid of $K_0$ dpa/s, the flux of atoms (atoms per unit area per second) from the precipitate surface was presented as $\xi K_0$, where $\xi$ is the irradiation re-solution constant evaluated in [15] as $\sim 10$ nm. Correspondingly, the volume $V$ growth rate of a spherical precipitate of radius $R$ was given by

$$c_p \frac{dV}{dt} = J_{dif} - 4\pi R^2 \xi K_0, \tag{1}$$

where $c_p$ is the concentration (dimensionless) of solute in the precipitates ($\approx 0.25$ for $Ni_3Al$ precipitates), neglecting the difference in atomic volumes between the precipitate phase and the matrix ($v \approx v_m$); the first term in the r.h.s., $J_{dif}$, is the diffusion flux of solute atoms into the precipitate, and the second term is the rate of precipitate volume change induced by ballistic re-solution.

A recent analysis of recoil dissolution of $\alpha'$-precipitates in Fe-18Cr alloys under ion irradiation [17], which represents the volumetric dissolution rate of precipitates by recoil as $(dV/dt)_{dis} = -4\pi R^2 \phi v$, where $\phi \approx f_m K_0$ is the flux of recoils from the cluster to the matrix, estimated the multiplication factor as $f_m \approx 1.17 \cdot 10^{21}$ m$^{-2}$. This gives $\xi/c_p \approx 15$ nm, and in the case $c_p = 0.25$ the re-solution constant can be estimated as $\xi \approx 4$ nm, which will be used in the present model.

A further improvement in the treatment of the dissolution rate in the NHM model was obtained by Brailsford [18] who assumed that ballistic re-solution occurs at a uniform rate in a shell of thickness $\lambda \sim 10$ nm (so called recoil distance) around the particle. This creates a source term $K_R$ in the diffusion equation for the solute concentration (dimensionless) $c(r)$ in this shell [19], which modifies the growth rate equation, Eq. (1), to the form

$$c_p \frac{dV}{dt} = 4\pi D R \left[ \bar{c} - c_s - \frac{K_R \lambda^2 (3R + 2\lambda)}{6DR} \right], \tag{2}$$

where $\bar{c}$ is the solute concentration at a large distance from the particle, $c_s$ is the interface solute concentration, $D$ is the irradiation-enhanced diffusion coefficient of solute atoms, and $K_R = 3R^2 \xi K_0 / [(R + \lambda)^3 - R^3]$.

Brailsford also noted that in the limit $R \ll \lambda$, the re-solution rate in Eq. (2) is consistently reduced to that in Eq. (1). For this reason, for the analysis of the nucleation kinetics of precipitates of a typical size $R \sim 1$ nm, Eq. (2) can be simplified as

$$c_p \frac{dV}{dt} \approx 4\pi DR \left(\bar{c} - c_s - \frac{\xi K_0}{D} R\right) = 4\pi DR \left(\bar{c} - c_{e,\infty} \exp\left(\frac{2\gamma v}{RkT}\right) - \frac{\xi K_0}{D} R\right), \quad (3)$$

where $c_{e,\infty}$ is equilibrium concentration near a flat interface (when $R \to \infty$), $\gamma$ is surface energy of the interface, and $c_s$ is calculated from the Gibbs-Kelvin equilibrium condition at the particle interface, $c_s(R) = c_{e,\infty} \exp(2\gamma v / RkT)$.

It is straight forward to show that more sophisticated models considering non-uniform distribution of ejected atoms in a shell of thickness $\lambda$ (e.g. [20]) can be similarly reduced to Eq. (3) under the above condition $R \ll \lambda$ (as shown, e.g., in Appendix A to [21]).

In terms of the supersaturation ratio defined as $S = \bar{c}/c_{e,\infty}$, the growth rate equation takes the form,

$$c_p \frac{dV}{dt} = c_p \frac{dx}{dt} v \approx 4\pi DR c_{e,\infty} \left(S - \exp\left(\frac{2\gamma v}{RkT}\right) - \frac{\xi K_0}{Dc_{e,\infty}} R\right), \quad (4)$$

where $x$ is the number of solute atoms of volume $v$ in the particle.

It should be noted that the use of the growth rate equation, controlled by diffusion in the matrix, is applicable only to incoherent particles. Indeed, the incoherent interface can be viewed as an atomically rough interface that can absorb and emit solute atoms (as well as other point defects). For this reason, the growth kinetics of incoherent precipitates, similarly to normal growth of crystals with atomically rough faces, occur by random attachment of atoms at the interface, which is controlled by the diffusion transport of solute atoms from the matrix to the interface.

However, in the case of a coherent interface, solute atoms can only be trapped (or *ad*sorbed) on an atomically smooth interface (rather than *ab*sorbed as in the case of a rough interface). Therefore, coherent interfaces form an essentially impenetrable barrier to atomic movement through them, and, for this reason, the growth of the coherent interface can be strongly suppressed [22]. In the absence of structural ledges (existing on semi-coherent interfaces), the growth of coherent particles, similarly to the growth of faceted crystals with atomically smooth interface [23, 24], is controlled by the formation of two-dimensional (2D) terraces as sources of growth layers. This requires modification of the traditional approach [15, 18] (applicable to incoherent particles), which will be discussed in Section 5.

### 3. Nucleation kinetics of precipitates

At first, for simplicity, nucleation kinetics of the single-component precipitate phase (corresponding to $c_p = 1$) will be analysed. According to classical nucleation theory [25–27], nucleation is represented by the translation of clusters (with size $x$ and number density $N_x$) in the phase space of cluster size $x$ (cf. also [28]). Correspondingly, the nucleation rate is calculated as a steady state flux $J_x$ of clusters between adjacent size classes in the phase space, which obeys the steady state condition, $dN_x/dt = J_{x-1} - J_x = 0$, or $J_x = \text{const}$, and takes the form,

$$J_x = \beta(x-1)s_{x-1}N_{x-1} - \alpha^{(th)}(x)s_x N_x, \quad (5)$$

where $\beta(x) = DS c_{e,\infty} R_x^{-1} v^{-1}$ is the arrival rate of monomers (solute atoms) by diffusion from the matrix to a particle ($x$ – mer) of radius $R_x = (3vx/4\pi)^{1/3}$ and surface area $s_x = 4\pi R_x^2$; and $\alpha^{(th)}(x)$ is the thermal loss rate, determined from the Gibbs-Kelvin equilibrium condition as

$$\alpha^{(th)}(x) = \beta(x) \exp\left(\frac{1}{kT} \frac{d\Delta G_0(x)}{dx}\right) = \beta(x) S^{-1} \exp\left(\frac{2\gamma v}{R_x kT}\right), \quad (6)$$

where $\Delta G_0(x)$ is the Gibbs free energy of the precipitate formation,

$$\Delta G_0(x) = \gamma(36\pi v^2)^{1/3} x^{2/3} - kTx\ln S. \tag{7}$$

Here, the contribution of elastic strain energy is neglected, corresponding to zero mismatch, $\varphi = 0$. For $\varphi \neq 0$, separate consideration will be given to incoherent particles (in Section 4) and coherent particles (in Section 5).

Under irradiation conditions, the steady state flux in phase space, in addition to thermal loss, must take into account ballistic re-solution and, accordingly, is modified to the form,

$$J_x = \beta(x-1)s_{x-1}N_{x-1} - [\alpha^{(th)}(x) + \alpha^{(irr)}(x)]s_x N_x, \tag{8}$$

$\alpha^{(irr)}(x) = \xi K_0 v^{-1}$ is the ballistic re-solution rate with the parameters $\xi$ and $K_0$ defined above in Eqs (2)–(4).

In the continuum approximation of classical nucleation theory, the cluster number density $N_x$ is transformed into the size distribution function $\rho(x)$, and the steady state flux, which determines the nucleation rate, takes the form,

$$J_x \approx \omega(x)\left\{-\frac{d\rho(x)}{dx} + \rho(x)S^{-1}\left[S - \exp\left(\frac{2\gamma v}{kTR_x}\right) - \frac{R_x \xi K_0}{Dc_{e,\infty}}\right]\right\} = \text{const}, \tag{9}$$

where $\omega(x) = \beta(x)s_x = 4\pi DR_x Sc_{e,\infty} v^{-1} = 3^{1/3}(4\pi/v)^{2/3} x^{1/3} DSc_{e,\infty}$.

The 'quasi-equilibrium' size distribution function $\rho_0(x)$, which is analogous to the equilibrium size distribution function in the absence of irradiation, is derived from the zero flux condition, $J_x = 0$, which represents the detailed balance between gaining and loss of monomers by clusters,

$$\beta(x-1)s_{x-1}\rho_0(x-1) - [\alpha^{(th)}(x) + \alpha^{(irr)}(x)]s_x \rho_0(x) = 0. \tag{10}$$

In the continuum approximation it takes the form,

$$-\frac{d\rho_0(x)}{dx} + \rho_0(x)S^{-1}\left[S - \exp\left(\frac{2\gamma v}{kTR_x}\right) - \frac{R_x \xi K_0}{Dc_{e,\infty}}\right] = 0, \tag{11}$$

or

$$\frac{d\ln\rho_0}{dx} = 1 - S^{-1}\frac{R_x \xi K_0}{Dc_{e,\infty}} - S^{-1}\exp\left(\frac{2\gamma v}{kTR_x}\right) = 1 - S^{-1}\frac{\xi K_0}{Dc_{e,\infty}}\left(\frac{3v}{4\pi}\right)^{\frac{1}{3}} x^{\frac{1}{3}} - S^{-1}\exp\left(ax^{-\frac{1}{3}}\right), \tag{12}$$

where $a = \left(\frac{4\pi}{3}\right)^{1/3}\frac{2\gamma v^{2/3}}{kT}$, whose solution is

$$\rho_0(x) = C\exp\left[x - \frac{3}{4}S^{-1}\frac{\xi K_0}{Dc_{e,\infty}}\left(\frac{3v}{4\pi}\right)^{\frac{1}{3}} x^{\frac{4}{3}} - S^{-1}\int \exp\left(ax^{-\frac{1}{3}}\right)dx\right] \equiv C\exp\left[-\frac{\Delta \bar{G}_0(x)}{kT}\right], \tag{13}$$

where $C$ is an (unknown) integration constant, and $\bar{G}_0(x)$ is a (non-thermodynamic) function that is a 'quasi-equilibrium' analogue of the free energy of cluster formation in the absence of irradiation $\Delta G_0(x)$ defined in Eq. (7).

The position of an extremum of $\rho_0(x)$, calculated using Eq. (13), coincides with that of $\Delta \bar{G}_0(x)$ and determines the critical size $x^*$ from the solution of the equation,

$$\rho_0^{-1}(x^*)\frac{d\rho_0(x^*)}{dx} = -\frac{d\bar{G}_0(x^*)}{kTdx} = 1 - S^{-1}\frac{\xi K_0}{Dc_{e,\infty}}\left(\frac{3v}{4\pi}\right)^{\frac{1}{3}} x^{*\frac{1}{3}} - S^{-1}\exp\left(ax^{*-\frac{1}{3}}\right) = 0, \tag{14}$$

which is consistent with the condition of zero growth rate of the critical nucleus arising from Eq. (4). Indeed, the growth kinetics of clusters can be described in the continuum approximation as

$$\dot{x} = s_x\big(\beta(x) - [\alpha^{(th)}(x) + \alpha^{(irr)}(x)]\big) = \beta(x)s_x\left[1 - \exp\left(\frac{1}{kT}\frac{d\Delta\bar{G}_0(x)}{dx}\right)\right], \tag{15}$$

which, after substituting the expression for $d\Delta\bar{G}_0(x)/dx$ from Eq. (14), coincides with Eq. (4) and turns to zero under the condition $d\Delta\bar{G}_0(x)/dx = 0$.

Using Eq. (13), one can represent Eq. (9) in the standard form of classical nucleation theory,

$$J_x = -\omega(x)\rho_0(x)\frac{d}{dx}\left(\frac{\rho(x)}{\rho_0(x)}\right) = \text{const}, \tag{16}$$

which allows applying the formal procedure of this theory, leading to the expression for the nucleation rate,

$$\dot{N} \approx \left(\int_0^\infty \frac{dx}{\omega(x)\rho_0(x)}\right)^{-1} \approx \omega^* \rho_0(x^*) Z, \tag{17}$$

where $x^*$ is determined by the extremum of $\rho_0(x)$ (as explained above), $\omega^* = \omega(x^*) = 3^{1/3}(4\pi/v)^{2/3}x^{*1/3}DSc_{e,\infty}$, and $Z$ is the Zeldovich factor.

In a small vicinity of the critical point $x = x^*$, Eq. (13) can be represented in the form

$$\rho_0(x) \approx \rho_0(x^*)\exp\left[-\frac{1}{2kT}\frac{d^2\Delta\bar{G}_0(x^*)}{dx^2}(x-x^*)^2\right], \tag{18}$$

where the second derivative of $\Delta\bar{G}_0(x)$ determines the Zeldovich factor,

$$Z = \left[-\frac{1}{2\pi kT}\frac{d^2\Delta\bar{G}_0(x^*)}{dx^2}\right]^{\frac{1}{2}} = \left\{\frac{1}{6\pi}\left[ax^{*-\frac{4}{3}} - S^{-1}\frac{\xi K_0}{Dc_{e,\infty}}\left(\frac{3v}{4\pi}\right)^{\frac{1}{3}}x^{*-\frac{2}{3}} - S^{-1}\frac{\xi K_0}{Dc_{e,\infty}}\left(\frac{3v}{4\pi}\right)^{\frac{1}{3}}ax^{*-1}\right]\right\}^{\frac{1}{2}}. \tag{19}$$

However, because of the unknown value of the integration constant $C$ in Eq. (13), $\rho_0(x^*)$ cannot be fully determined in the continuum approximation. Conversely, it can be calculated directly from the discrete Eq. (10) using Fuchs' algorithm [29] developed for non-equilibrium conditions (see also [30]), which leads to the recurrence relationship,

$$\frac{\rho_0(x-1)}{\rho_0(x)} \approx \frac{\alpha^{(th)}(x)+\alpha^{(irr)}(x)}{\beta(x)} = S^{-1}\left[\exp\left(\frac{2\gamma v}{R_x kT}\right) + \frac{\xi K_0 R_x}{Dc_{e,\infty}}\right]. \tag{20}$$

and, consequently,

$$\frac{N_1}{\rho_0(x)} = \frac{N_1}{\rho_0(2)}\frac{\rho_0(2)}{\rho_0(3)}\cdots\frac{\rho_0(x-1)}{\rho_0(x)} \approx \prod_{i=2}^{x} S^{-1}\left[\exp\left(\frac{2\gamma v}{R_i kT}\right) + \frac{\xi K_0 R_i}{Dc_{e,\infty}}\right], \tag{21}$$

or

$$\ln\frac{N_1}{\rho_0(x)} \approx \sum_{i=2}^{x}\ln\left\{S^{-1}\left[\exp\left(\frac{2\gamma v}{kTR_i}\right) + \frac{\xi K_0 R_i}{Dc_{e,\infty}}\right]\right\}, \tag{22}$$

where $N_1 = \bar{c}/v = Sc_{e,\infty}/v$ is the number density of monomers in the matrix. This algorithm was employed to analyse void nucleation under irradiation conditions in [31, 32] and further developed in the author's work [33].

Similar to the calculations in [33], Eq. (22) can be represented as

$$\ln\frac{N_1}{\rho_0(x)} = \sum_{k=2}^{n}\ln\left\{S^{-1}\exp\left(\frac{2\gamma v}{kTR_k}\right)\left[1 + \frac{\xi K_0}{Dc_{e,\infty}}R_k\exp\left(-\frac{2\gamma v}{kTR_k}\right)\right]\right\} = \sum_{k=2}^{n}\left\{\ln\left[S^{-1}\exp\left(\frac{2\gamma v}{kTR_k}\right)\right] + \ln\left[1 + \frac{\xi K_0}{Dc_{e,\infty}}R_k\exp\left(-\frac{2\gamma v}{kTR_k}\right)\right]\right\}, \tag{23}$$

and then, by passing from summation to integration, is transformed to the form,

$$\ln \frac{N_1}{\rho_0(x)} \approx \left[\frac{3\gamma v}{kT}\left(\frac{4\pi}{3v}\right)^{\frac{1}{3}} x^{\frac{2}{3}} - x \ln S\right] + \int_0^x \ln\left[1 + \frac{\xi K_0}{Dc_{e,\infty}}\left(\frac{3v}{4\pi}\right)^{1/3} y^{\frac{1}{3}} \exp\left(-ay^{-\frac{1}{3}}\right)\right] dy, \quad (24)$$

or

$$\rho_0(x) \approx N_1 \exp\left[x \ln S - \frac{3}{2} a x^{\frac{2}{3}} - I(x)\right], \quad (25)$$

where

$$I(x) = \int_0^x \ln\left[1 + \frac{\xi K_0}{Dc_{e,\infty}}\left(\frac{3v}{4\pi}\right)^{1/3} y^{\frac{1}{3}} \exp\left(-ay^{-\frac{1}{3}}\right)\right] dy. \quad (26)$$

It is straightforward to show that Eq. (25) correctly determines the position $x = x^*$ of the critical point, found from Eq. (14), and the value of the Zeldovich factor, found from Eq. (19)). Therefore, although the functional dependence of $\rho_0$ on $x$ differs from Eq. (13), it correctly describes the position and form of $\rho_0(x)$ in a small vicinity of its extremum (located at the critical point). This suggests a satisfactory accuracy of this solution at the critical point, which gives

$$\rho_0(x^*) \approx S c_{e,\infty} \exp\left[x^* \ln S - \frac{3}{2} a x^{*\frac{2}{3}} - I(x^*)\right]. \quad (27)$$

This result requires some additional justification, which is presented in Appendix A.

Therefore, to evaluate the nucleation rate, which is obtained by substituting Eqs (19) and (27) into Eq. (17),

$$\dot{N} \approx$$

$$3^{1/3}(4\pi)^{\frac{2}{3}} DS^2 c_{e,\infty}^2 v^{-\frac{5}{3}} x^{*\frac{1}{3}} \left\{\frac{1}{6\pi}\left[ax^{*-\frac{4}{3}} - S^{-1}\frac{\xi K_0}{Dc_{e,\infty}}\left(\frac{3v}{4\pi}\right)^{\frac{1}{3}} x^{*-\frac{2}{3}} - \right.\right.$$

$$\left.\left. S^{-1}\frac{\xi K_0}{Dc_{e,\infty}}\left(\frac{3v}{4\pi}\right)^{\frac{1}{3}} a x^{*-1}\right]\right\}^{\frac{1}{2}} \exp\left[x^* \ln S - \frac{3}{2} a x^{*\frac{2}{3}} - I(x^*)\right], \quad (28)$$

one should determine the critical nucleus size $x^*$ from Eq. (14).

In some vicinity of the saturation point where the critical size is relatively large, so that $2\gamma v/kTR^* \ll 1$, Eq. (14) can be simplified to the form,

$$S - \left(1 + \frac{2\gamma v}{kTR^*}\right) - \frac{R^*\xi K_0}{Dc_{e,\infty}} = 0, \quad (29)$$

or

$$R^{*2} - R^*(S-1)\frac{Dc_{e,\infty}}{\xi K_0} + \frac{2\gamma v}{kT}\frac{Dc_{e,\infty}}{\xi K_0} = 0, \quad (30)$$

which has two roots,

$$R^*_{1,2} = \frac{Dc_{e,\infty}(S-1)}{2\xi K_0}\left\{1 \mp \left[1 - \frac{8\gamma v \xi K_0}{kTDc_{e,\infty}(S-1)^2}\right]^{1/2}\right\}, \quad (31)$$

existing only if the discriminant is positive.

The first root, $R^*_1$, provides a maximum (or an unstable extremum) for $\bar{G}_0(x)$ and thus determines the radius of the critical cluster, whereas the second root, $R^*_2$, providing a minimum (a stable extremum) for $\bar{G}_0(x)$, determines the maximum radius of the particle. In this case, the growth of the nucleated particle, governed by Eq. (4), will occur until the stable extremum $R^*_2$ is reached.

Accordingly, the condition of a positive determinant in Eq. (31) defines the threshold at which nucleation becomes possible, which is found from the equation

$$\frac{8\gamma v \xi K_0}{kTDc_{e,\infty}(S-1)^2} = 1, \tag{32}$$

the solution of which determines the threshold value of the supersaturation ratio,

$$\hat{S} = 1 + \left(\frac{8\gamma v \xi K_0}{kTDc_{e,\infty}}\right)^{\frac{1}{2}}, \tag{33}$$

below which particle nucleation is unfeasible. In particular, this threshold effect may explain the suppression of new phase formation observed in experiments (see Section 1), if the supersaturation ratio in the solid solution is below the threshold value, $1 \leq S < \hat{S}$.

At this threshold, the critical radius has a finite value

$$\hat{R} = \left(\frac{2\gamma v D c_{e,\infty}}{kT\xi K_0}\right)^{\frac{1}{2}} = \frac{4\gamma v}{kT(\hat{S}-1)}, \tag{34}$$

which should satisfy the above assumption $2\gamma v/kT\hat{R} \ll 1$ to make these estimates accurate.

In a more general case (when $2\gamma v/kT\hat{R} \ll 1$ is not satisfied), the threshold radius can be determined from the condition of merging two extrema of the function $\Delta \bar{G}_0(x)$, which has the form $d^2\Delta \bar{G}_0(x)/dx^2 = 0$, and can be calculated by differentiating Eq. (14), giving

$$\exp\left(\frac{2\gamma v}{RkT}\right) = R^2 \frac{kT\xi K_0}{2\gamma v D c_{e,\infty}}, \tag{35}$$

which solution $\hat{R}$ consistently converges to Eq. (34) in the limit $2\gamma v/kT\hat{R} \ll 1$.

The threshold value of the supersaturation ratio in the general case is calculated by substituting Eq. (34) into Eq. (14), giving

$$\hat{S} = \hat{R}\frac{\xi K_0}{Dc_{e,\infty}}\left(\frac{kT}{2\gamma v}\hat{R} + 1\right), \tag{36}$$

which consistently converges to Eq. (33) in the limit $2\gamma v/kT\hat{R} \ll 1$.

As seen from Eq. (34), the threshold radius $\hat{R}$ has a finite value, whereas the critical radius $R^*$ (the first root of Eq. (31)) increases smoothly, starting from this value, $R^* \geq \hat{R}$, with increasing supersaturation, $S \geq \hat{S}$ (in contrast to $R^* \to \infty$ near the saturation point in the absence of ballistic re-solution). For this reason, in a certain vicinity of the threshold supersaturation, the nucleation rate can be well characterised by its value at the threshold.

For Ni$_3$Al precipitates with $c_p = 0.25$ in Ni-Al alloys studied in the ion bombardment tests [15], the above theory needs to be somewhat modified, as shown in Appendix B. For these precipitates, the surface energy of the interface, according to first-principles studies [34], drops rapidly with increasing temperature and becomes practically constant, $\gamma \approx 10^{-2}$ J·m$^{-2}$, for temperatures above 600 K, which is in very good agreement with previous fits to experimentally measured coarsening rates [35]. For the conditions of the ion bombardment tests [15] at $T \approx 550$°C with the parameters $K_0 \approx 10^{-2}$ s$^{-1}$, $\xi \approx$ 4 nm (discussed above in Section 2), $D \approx 6 \cdot 10^{-18}$ m$^2$/s (estimated in [15]), $c_{e,\infty} \approx 0.1$ (estimated from the binary Ni-Al equilibrium phase diagram), and $v \approx 1.23 \cdot 10^{-29}$ m$^3$, the threshold supersaturation ratio can be estimated from Eq. (B.13) as $\hat{S} \approx 1.3$. This corresponds to an increase in the limit of non-equilibrium solubility compared to equilibrium solubility, i.e. to a decrease in the stability range of the new phase. The threshold radius can be estimated from Eq. (B.14) as $\hat{R} \approx$ 0.8 nm, which well satisfies the above condition $2\gamma v/kT\hat{R} \approx 0.03 \ll 1$, and thus can be considered a good estimate for the test conditions [15] (which will be used in the further analysis of these tests in Section 6).

In the case of high dislocation density, $\rho_d \sim 10^{15}$ m$^{-2}$, considered in [15], which provides dominant loss of point defects to sinks, the radiation enhanced diffusion coefficient was estimated as $D \approx D_v c_v \approx K_0/\rho_d$, where $D_{v(i)}$ and $c_{v(i)}$ are diffusivity and concentration of vacancies (interstitials), respectively. In this case, the threshold supersaturation and radius, calculated in Eqs (B.13) and (B.14) (modifying Eqs (33) and (34)), do not depend on the damage rate, $\hat{S}' = \hat{S}^{(1-3c_{e,\infty})} \approx 1 + \left(\frac{8\gamma v \xi \rho_d}{kT c_{e,\infty} c_p}\right)^{1/2}$, and $\hat{R} \approx \left(\frac{4\gamma v c_{e,\infty}}{kT \xi c_p \rho_d}\right)^{1/2}$, and thus the above estimates of the threshold values can also be applied to the conditions of neutron irradiation with $K_0 \approx 10^{-6}$ s$^{-1}$.

The expression for $I(x^*)$, derived from Eq. (B.12) (modifying Eq. (26)), can be simplified in the vicinity of the threshold supersaturation $\hat{S}$ (where $R^* \approx \hat{R}$, or $x^* \approx \hat{x}$), leading to

$$I'(x^*) \approx \int_0^{x^*} \frac{\xi K_0}{D c_{e,\infty}} \left(\frac{3v}{4\pi c_p}\right)^{1/3} y^{\frac{1}{3}} \exp\left(-a' c_p^{2/3} y^{-\frac{1}{3}}\right) dy, \quad (37)$$

since $\frac{\xi K_0}{D c_{e,\infty}} \left(\frac{3v}{4\pi c_p}\right)^{1/3} x^{*\frac{1}{3}} \exp\left(-a' c_p^{2/3} x^{*-\frac{1}{3}}\right) \ll 1$, where $a' = a c_p^{-1/3}$ This expression can be integrated to obtain

$$I'(x^*) \approx \frac{\xi K_0}{8 D c_{e,\infty}} \left(\frac{3v}{4\pi c_p}\right)^{1/3} \left[x^{*\frac{1}{3}}\left(6x^* - 2a' c_p^{2/3} x^{*\frac{2}{3}} + a'^2 c_p^{4/3} x^{*\frac{1}{3}} - c_p^2 a'^3\right) \exp\left(-a' c_p^{2/3} x^{*-\frac{1}{3}}\right) -$$

$$a'^4 c_p^{8/3} \mathrm{Ei}\left(-a' c_p^{2/3} x^{*-\frac{1}{3}}\right)\right], \quad (38)$$

(where $\mathrm{Ei}(z) = -\int_{-z}^{\infty} t^{-1} \exp(-t) dt$ is exponential integral), which reflects the effect of suppression of the nucleation rate by ballistic re-solution.

Under the condition $\varepsilon \equiv a' c_p^{2/3} \hat{x}^{-\frac{1}{3}} = 2\gamma v c_p^{1/3}/kT\hat{R} \ll 1$, which is valid for the tests [15] (as estimated above), Eq. (38) can be further simplified as

$$I'(x^*) \approx \frac{2\pi \gamma^2 v D c_{e,\infty}}{3 c_p^3 \xi K_0 (kT)^2} [(6 - 2\varepsilon + \varepsilon^2 - \varepsilon^3) \exp(-\varepsilon) - \varepsilon^4 \mathrm{Ei}(-\varepsilon)] \approx \frac{4\pi \gamma^2 v D c_{e,\infty}}{c_p^3 \xi K_0 (kT)^2}, \quad (39)$$

given $\mathrm{Ei}(-\varepsilon) \approx 0.577 + \ln \varepsilon$ in the considered limit $\varepsilon \ll 1$.

In the same limit, the 'thermal' term in the exponent of Eq. (B.11) takes the form (at $S' \approx \hat{S}'$),

$$I'_{th}(x^*) = x^* \ln S' - \frac{3}{2} \frac{a'}{c_p^{1/3}} x^{*\frac{2}{3}} \approx -\frac{8\pi \gamma^2 v D c_{e,\infty}}{3 c_p^3 \xi K_0 (kT)^2}, \quad (40)$$

and, therefore, the relative contribution of the irradiation term can be estimated as

$$I'(x^*)/I'_{th}(x^*) \approx -1.5, \quad (41)$$

which indicates a significant suppression of the nucleation rate, only partially compensated by the enhancement of diffusion in the pre-exponential (kinetic) part of the nucleation rate.

Therefore, the exponential factor in the nucleation rate at threshold supersaturation can be estimated as

$$\dot{N} \propto \exp\left[\hat{x} \ln S' - \frac{3}{2} \frac{a'}{c_p^{1/3}} \hat{x}^{\frac{2}{3}} - I'(\hat{x})\right] \approx \exp\left[-\frac{20\pi \gamma^2 v D c_{e,\infty}}{3 c_p^3 \xi K_0 (kT)^2}\right] \approx \exp\left[-\frac{20\pi \gamma^2 v c_{e,\infty}}{3 c_p^3 \xi \rho_d (kT)^2}\right], \quad (42)$$

which demonstrates a strong (parabolic) dependence of the nucleation rate on the interfacial surface energy $\gamma$, which is relatively low ($\approx 10^{-2}$ J·m$^{-2}$) for Ni$_3$Al precipitates in Ni-Al alloys, but may increase by one order of magnitude in other binary systems. In such systems, the absolute value of the exponent increases by two orders of magnitude, which can lead to much stronger suppression of the nucleation rate near the supersaturation threshold.

## 4. Nucleation of incoherent precipitates

As emphasised in Section 1, ballistic effects act simultaneously with the effects of the excess of point defects formed under steady state irradiation conditions and should therefore be taken into account in a coordinated manner. Therefore, the influence of the volume mismatch strain on the nucleation kinetics of incoherent precipitates, which is affected by an excess of point defects in the irradiated matrix [8], should be additionally incorporated in the above developed model.

For this purpose, the thermal loss rate of solute atoms from an incoherent precipitate should be re-evaluated in Eq. (5), taking into account the contribution of strain energy to the free energy of precipitate formation, $\Delta G_{el} = 6\mu \left(\frac{3K}{3K+4\mu}\right)\delta^2 V = 6\mu' \delta^2 V$, where $\mu$ is the shear modulus of the matrix, $K$ is the bulk modulus of the precipitate, $\mu' \equiv \mu\left(\frac{3K}{3K+4\mu}\right)$, and $\delta \approx (V_p - V_m)/3V_p \ll 1$ is the transformation strain due to forming the particle of a volume $V_p = xv$, consisting of $x$ solute atoms and placed in a spherical cavity of a volume $V_m = xv_m$; $v$ and $v_m$ are the atomic volumes in the particle and in the matrix respectively (cf. [36]).

As shown in [8] (and refined in [9, 10]), in the case of single-component precipitates in the matrix oversaturated with vacancies, the formation free energy takes the form,

$$\Delta G_0(x,n) = -kTx \ln S - kTn \ln S_v + 4\pi\gamma \left(\frac{3}{4\pi}\frac{v}{1+\varphi}\right)^{\frac{2}{3}} (x+n)^{\frac{2}{3}} + \frac{2}{3}\mu'v\left(\frac{1}{1+\varphi}\right)^2 x\left(\varphi - \frac{n}{x}\right)^2, \quad (43)$$

where $S_v$ is the vacancy supersaturation ratio in the matrix, $n$ is the number of vacancies absorbed at the interface of the particle consisting of $x$ solute atoms that increase the volume of the enclosing cavity from $V_m$ to $V_m' = (x+n)v_m$ and decrease the transformation strain from $\delta$ to $\delta' = (V - V_m')/3V \approx \frac{1}{3}\left(\varphi - \frac{n}{x}\right)\left(\frac{1}{1+\varphi}\right)$, where $\varphi = (v - v_m)/v_m \ll 1$ is the volume mismatch parameter. The two last terms in the r.h.s. of Eq. (43) represent, respectively, the surface energy $\gamma s_{x,n}$, where $s_{x,n} = 4\pi R^2(x,n)$ is the surface area of the interface of radius $R(n,x) = [3(x+n)v_m/4\pi]^{1/3}$, and the elastic energy $\Delta G_{el}$ associated with the transformation strain $\delta'$.

Therefore, the thermal loss rate of solute atoms from a particle (cluster) consisting of $x$ solute atoms and with $n$ vacancies absorbed at the interface can be calculated using the Gibbs-Kelvin equilibrium condition as

$$\alpha_x^{(th)}(x,n) = \beta_x(x,n) \exp\left(\frac{1}{kT}\frac{d\Delta G_0(x,n)}{dx}\right) = \beta_x(x,n) \exp\left[-\ln S + \frac{8\pi\gamma}{3kT}\left(\frac{3}{4\pi}\frac{v}{1+\varphi}\right)^{\frac{2}{3}} x^{-\frac{1}{3}}\left(1+\frac{n}{x}\right)^{-\frac{1}{3}} + \right.$$

$$\left. \frac{2\mu'v}{3kT}\left(\frac{1}{1+\varphi}\right)^2 \left[\varphi^2 - \left(\frac{n}{x}\right)^2\right]\right], \quad (44)$$

where $\beta_x(n,x) = DSc_{e,\infty}R(n,x)^{-1}v^{-1}$ is the arrival rate of solute atoms to the cluster $(x,n)$, while the thermal loss rate of vacancies is

$$\alpha_v^{(th)}(x,n) = \beta_v(x,n) \exp\left(\frac{1}{kT}\frac{d\Delta G_0(x,n)}{dn}\right) = \beta_v(x,n) \exp\left[-\ln S_v + \frac{8\pi\gamma}{3kT}\left(\frac{3}{4\pi}\frac{v}{1+\varphi}\right)^{\frac{2}{3}} x^{-\frac{1}{3}}\left(1+\frac{n}{x}\right)^{-\frac{1}{3}} - \frac{4\mu'v}{3kT}\left(\frac{1}{1+\varphi}\right)^2\left(\varphi-\frac{n}{x}\right)\right], \tag{45}$$

where $\beta_v(x,n) = D_v c_v R(x,n)^{-1} v^{-1}$ is the arrival rate of vacancies.

For irradiation conditions, in order to calculate the steady state flux in the phase space of cluster sizes $(x,n)$ in the direction $x$, Eq. (44) should be substituted into the expression (that modifies Eq. (5)),

$$J_x = \beta_x(x-1,n) s_{x-1,n}\rho(x-1,n) - \left[\alpha_x^{(th)}(x,n) + \alpha_x^{(irr)}\right] s_{n,x}\rho(x,n), \tag{46}$$

where $\alpha_x^{(irr)} = \xi K_0 v^{-1}$, while the steady state flux in the direction $n$ is calculated by substituting Eq. (45) into the expression

$$J_n = \beta_v(x,n-1) s_{x,n-1}\rho(x,n-1) - \alpha_v^{(th)}(x,n) s_{x,n}\rho(x,n) - \beta_i(x,n) s_{x,n}\rho(x,n), \tag{47}$$

where $\beta_i(x,n) = D_i c_i R(x,n)^{-1} v^{-1}$ is the arrival rate of interstitials. The probability of interstitial loss is so small as to be negligible.

Accordingly, the quasi-equilibrium size distribution function $\rho_0(x,n)$ is derived from the condition of zero fluxes, $J_n = 0$ and $J_x = 0$, which, similarly to the derivation of Eq. (12), leads to

$$-\frac{d(\ln\rho_0(x,n))}{dn} = 1 - \frac{\beta_i}{\beta_v} - S_v^{-1} \exp\left[\frac{8\pi\gamma}{3kT}\left(\frac{3}{4\pi}\frac{v}{1+\varphi}\right)^{\frac{2}{3}} x^{-\frac{1}{3}}\left(1+\frac{n}{x}\right)^{-\frac{1}{3}} - \frac{4\mu'v}{3kT}\left(\frac{1}{1+\varphi}\right)^2\left(\varphi-\frac{n}{x}\right)\right], \tag{48}$$

and

$$-\frac{d(\ln\rho_0(x,n))}{dx} = 1 - \frac{\xi K_0}{DSc_{e,\infty}}\left[\frac{3v}{4\pi(1+\varphi)}\right]^{\frac{1}{3}} x^{\frac{1}{3}}\left(1+\frac{n}{x}\right)^{\frac{1}{3}} - S^{-1} \exp\left\{\frac{8\pi\gamma}{3kT}\left(\frac{3}{4\pi}\frac{v}{1+\varphi}\right)^{\frac{2}{3}} x^{-\frac{1}{3}}\left(1+\frac{n}{x}\right)^{-\frac{1}{3}} + \frac{2\mu'v}{3kT}\left(\frac{1}{1+\varphi}\right)^2\left[\varphi^2 - \left(\frac{n}{x}\right)^2\right]\right\}. \tag{49}$$

The position of an extremum of $\rho_0(x,n)$ determines the critical cluster parameters $x^*$ and $n^*$, calculated by equating Eqs (48) and (49) to zero,

$$ax^{*-\frac{1}{3}}\left(\frac{1}{1+\varphi}\right)^{\frac{2}{3}}\left(1+\frac{n^*}{x^*}\right)^{-\frac{1}{3}} - \frac{4\mu'v}{3kT}\left(\frac{1}{1+\varphi}\right)^2\left(\varphi-\frac{n^*}{x^*}\right) = \ln\tilde{S}_v, \tag{50}$$

and

$$ax^{*-\frac{1}{3}}\left(\frac{1}{1+\varphi}\right)^{\frac{2}{3}}\left(1+\frac{n^*}{x^*}\right)^{-\frac{1}{3}} + \frac{2\mu'v}{3kT}\left(\frac{1}{1+\varphi}\right)^2\left[\varphi^2 - \left(\frac{n^*}{x^*}\right)^2\right] = \ln\tilde{S}, \tag{51}$$

where $\tilde{S}_v = S_v\left(1-\frac{\beta_i}{\beta_v}\right)$ and $\tilde{S} = S\left[1 - \frac{\xi K_0}{DSc_{e,\infty}}\left[\frac{3v}{4\pi(1+\varphi)}\right]^{\frac{1}{3}} x^{*\frac{1}{3}}\left(1+\frac{n^*}{x^*}\right)^{\frac{1}{3}}\right] \approx \left[S - \frac{\xi K_0}{Dc_{e,\infty}}\left[\frac{3v}{4\pi(1+\varphi)}\right]^{\frac{1}{3}} x^{*\frac{1}{3}}\right]$, under the assumption $n^*/x^* \ll 1$ (which will be confirmed below).

Superposition of Eqs (50) and (51) gives

$$\frac{2\mu v}{3kT}\left(\frac{1}{1+\varphi}\right)^2\left[\varphi^2 - \left(\frac{n^*}{x^*}\right)^2\right] + \frac{4\mu'v}{3kT}\left(\frac{1}{1+\varphi}\right)^2\left(\varphi-\frac{n^*}{x^*}\right) = \ln\left(\frac{\tilde{S}}{\tilde{S}_v}\right), \tag{52}$$

or,

$$\frac{n^*}{x^*} = (1+\varphi)\left(1 - \frac{3kT}{2\mu' v}\ln\frac{\tilde{S}}{\tilde{S}_v}\right)^{1/2} - 1, \tag{53}$$

which confirms the above assumption $n^*/x^* \ll 1$ (given $\varphi \ll 1$).

Since $kT/\mu' v \sim 10^{-2} \ll 1$, the substitution of Eq. (53) into Eq. (50) gives

$$x^* \approx \frac{32\pi}{3}\left(\frac{\gamma}{kT}\right)^3 \frac{v^2}{\left[\ln\tilde{S} + \varphi\ln\tilde{S}_v + \frac{3}{8}\frac{kT}{\mu' v}\left(\ln\frac{\tilde{S}}{\tilde{S}_v}\right)^2\right]^3 \left(1 - \frac{3kT}{4\mu' v}\ln\frac{\tilde{S}}{\tilde{S}_v}\right)}. \tag{54}$$

Correspondingly, the expression for the formation free energy of the critical nucleus takes the form,

$$\Delta G_0^* \approx \frac{16\pi}{3}\frac{\gamma^3 v^2}{(kT)^2} \frac{1}{\left[\ln\tilde{S} + \varphi\ln\tilde{S}_v + \frac{3}{8}\frac{kT}{\mu' v}\left(\ln\frac{\tilde{S}}{\tilde{S}_v}\right)^2\right]^3}\left[\ln\tilde{S} + \varphi\ln\tilde{S}_v + \frac{3kT}{2\mu' v}\ln\frac{\tilde{S}}{\tilde{S}_v}\ln\tilde{S}_v + \frac{15}{8}\frac{kT}{\mu' v}\left(\ln\frac{\tilde{S}}{\tilde{S}_v}\right)^2\right], \tag{55}$$

which shows that the effect of elastic strain energy is negligible (due to compensation by point defects) and that the critical supersaturation shifts to $\tilde{S}^* \approx \tilde{S}_v^{-\varphi}$.

In the zero-order approximation for the small parameter $kT/\mu' v \approx 10^{-2}$, Eq. (54) can be simplified as

$$ax^{*-\frac{1}{3}} \approx (\ln\tilde{S} + \varphi\ln\tilde{S}_v), \tag{56}$$

or

$$\exp\left(ax^{*-\frac{1}{3}}\right) \approx \tilde{S}_v^{\varphi}\tilde{S} \approx \tilde{S}_v^{\varphi}\left[S - \frac{\xi K_0}{Dc_{e,\infty}}\left(\frac{3v}{4\pi}\right)^{\frac{1}{3}}x^{*\frac{1}{3}}\right], \tag{57}$$

which modifies Eq. (29) to the form,

$$S - \tilde{S}_v^{-\varphi}\left(1 + \frac{2\gamma v}{kTR^*}\right) - \frac{R^*\xi K_0}{Dc_{e,\infty}} \approx 0, \tag{58}$$

or

$$R^{*2} - R^*\left(S - \tilde{S}_v^{-\varphi}\right)\frac{Dc_{e,\infty}}{\xi K_0} + \frac{2\gamma v}{kT}\frac{Dc_{e,\infty}}{\xi K_0}\tilde{S}_v^{-\varphi} \approx 0, \tag{59}$$

whose solutions are

$$R_{1,2}^* = \frac{D\left(S - \tilde{S}_v^{-\varphi}\right)c_{e,\infty}}{2\xi K_0} \mp \left[\left(\frac{Dc_{e,\infty}}{2\xi K_0}\right)^2\left(S - \tilde{S}_v^{-\varphi}\right)^2 - \frac{2D\gamma vc_{e,\infty}}{\xi K_0 kT}\tilde{S}_v^{-\varphi}\right]^{1/2}. \tag{60}$$

Therefore, the threshold supersaturation ratio, determined from the condition of zero discriminant in Eq. (60), takes the form

$$\hat{S} \approx \tilde{S}_v^{-\varphi} + \left(\frac{8\gamma v\xi K_0}{DkTc_{e,\infty}}\tilde{S}_v^{-\varphi}\right)^{1/2}, \tag{61}$$

which consistently converges to Eq. (33) at $\varphi = 0$, and to the result of the Maydet-Russel model [12] for the critical supersaturation ratio, $S^* \approx \tilde{S}_v^{-\varphi}$, obtained in neglect of the ballistic re-solution effect (i.e. in the limit $\xi \to 0$).

Consequently, the threshold radius is calculated as

$$\hat{R} \approx \left(\frac{2\gamma vDc_{e,\infty}}{\xi K_0 kT}\tilde{S}_v^{-\varphi}\right)^{1/2}. \tag{62}$$

As noted in Section 1, the effect of excess vacancies is substantially reduced by interstitials, since the factor $(1 - \beta_i/\beta_v)$ under irradiation is relatively small (typically a few percent), which at relatively high temperatures (when $\tilde{S}_v^{-\varphi} \to 1$) can significantly reduce the influence of point defects on the threshold supersaturation (given $\varphi \ll 1$). This allows concluding that at high irradiation temperatures (when $\tilde{S}_v$ is small), ballistic effects can significantly impact the stability of the new incoherent phase.

Conversely, at lower temperatures, when $\tilde{S}_v$ increases considerably, the influence of point defects can dominate, although not as strongly as assumed by Russel [13] when comparing the nucleation of graphite and cementite phases in irradiated steels. He noted that a strong effect only arises in the case of $\varphi \approx 1$, which he attributed to the graphite phase. However, this is beyond the applicability of the theory (assuming $\varphi \ll 1$), and must be recalculated using the redefined value $\tilde{\varphi} = \varphi - 1 \ll 1$ (as explained in Appendix to [9, 10]), which leads to a much weaker effect.

## 5. Nucleation of coherent precipitates

As discussed in Section 2, in the case of coherent particles, point defects can only be trapped (or *ad*sorbed) on a smooth interface rather than *ab*sorbed (as in the case of incoherent particles with a rough interface). For this reason, the elastic strain energy is not compensated by point defects (as in the case of incoherent particles) and must be taken into account in the Gibbs free energy of spherical nucleus formation, which in the case of single-component precipitates ($c_p = 1$) takes the form,

$$\Delta G_0(x) = \gamma(36\pi v^2)^{1/3} x^{2/3} - kTx\ln S + 6\mu'\delta^2 xv, \tag{63}$$

which affects the stability of the new phase, shifting the critical supersaturation to $\ln S^* = 6\mu'\delta^2 v/kT$.

The contribution of excess point defects leads to a decrease in the interface surface energy, characterised in this case by the effective value $\gamma_{eff}$, whose dependence on irradiation conditions was calculated in the author's work [38],

$$\gamma_{eff} = \gamma + \frac{kT}{(36\pi v^2)^{\frac{1}{3}}} (9\pi)^{\frac{1}{3}} \left\{ \theta_v \left[ \ln\left(\frac{\theta_v}{\theta_v^{(0)}}\right) - \ln S_v \right] + \theta_i \left[ \ln\left(\frac{\theta_i}{\theta_i^{(0)}}\right) - \ln S_i \right] \right\}, \tag{64}$$

where $\theta_v$ and $\theta_i$ are steady state surface concentrations of adsorbed vacancies and interstitials, respectively, which are determined by their mutual recombination at the interface; $\theta_v^{(0)}$ and $\theta_i^{(0)}$ are their thermal concentrations; $S_i$ is the interstitial supersaturation ratio in the matrix. It was shown that under typical reactor irradiation conditions, the correction to $\gamma$ is negative and can significantly reduce its effective value $\gamma_{eff}$.

Consequently, Eq. (63) can be represented in the form,

$$\Delta G_0(x) = \gamma_{eff}(36\pi v^2)^{1/3} x^{2/3} - kTx\ln\tilde{S}, \tag{65}$$

where $S_{eff} = S\exp(-6\mu'\delta^2 v/kT)$, which can be obtained from Eq. (7) by formal substituting $\gamma \to \gamma_{eff}$ and $S \to S_{eff}$. Accordingly, all the results of Section 3, obtained by taking ballistic re-solution into account, can be recalculated using these substitutions. In particular, this will modify the threshold value of the supersaturation ratio, Eq. (33), which accordingly takes the form

$$\hat{S} \exp\left(-\frac{6\mu'\delta^2 v\mu}{kT}\right) = 1 + \left(\frac{8\gamma_{eff}v\xi K_0}{kTDc_{e,\infty}}\right)^{\frac{1}{2}}, \tag{66}$$

and demonstrates the effect of elastic strain energy on the stability of the new phase, shifting it to higher supersaturations. However, for Ni$_3$Al precipitates in Ni-Al alloys (considered below in Section

6) with $\mu \approx 75$ GPa, $K \approx 160$ GPa and the lattice misfit $\delta \approx 10^{-3}$ [39], this shift is extremely small, $6\mu' \delta^2 v/kT \sim 10^{-4} - 10^{-3}$, and therefore can be neglected compared to the shift caused by ballistic re-solution.

The nucleation rate, described by Eq. (28) with the modified parameters $S_{eff}$ and $\gamma_{eff}$ requires the additional modification of the kinetic part of the nucleation rate. In this case, the growth rate equation takes the form,

$$\frac{dx}{dt} = \tilde{\beta}(x) s_x [1 - S_{eff}^{-1} \exp(\tilde{a}x)] - \left(\frac{36\pi}{v}\right)^{1/3} x^{2/3} \xi K_0, \tag{67}$$

where $\tilde{a} = \left(\frac{4\pi}{3}\right)^{1/3} \frac{2\gamma_{eff} v^{2/3}}{kT}$, and the arrival rate $\tilde{\beta}(x)$ is determined by the relatively slow process of nucleation of two-dimensional terraces of monoatomic height on an atomically smooth interface (see Section 2). As shown in the author's work [40], this growth mechanism leads to a suppression of the nucleation rate due to a strong reduction in the parameter $\omega(x^*) = \tilde{\beta}(x^*) s_{x^*}$ in the nucleation rate, Eq. (17), by an exponential factor, which increases the nucleation barrier.

On the other hand, this decrease in the nucleation rate can be effectively counterbalanced by a decrease in the effective interface energy $\gamma_{eff}$, which lowers the nucleation barrier $\Delta G_0(x^*)$, as shown in [38]. Ultimately, competition between these effects can lead to substantial uncertainty in the value of the nucleation barrier, which for practical applications can be determined by fitting the calculation results (with $\gamma$ adjusted) to experimental data.

This approach is similar to earlier considerations of unirradiated materials, e.g. by Hyland [41] (in application to nucleation of coherent $Al_3Sc$ precipitates in a dilute Al-Sc alloy), who adjusted the average value of the interface energy $\gamma$ to a significantly higher value compared to existing experimental data and atomistic calculations for this quantity, in order to achieve reasonable agreement between the simplified nucleation model (neglecting the above discussed increase in the nucleation barrier for coherent precipitates) and his measurements of nucleation kinetics (cf. [40]).

## 6. Results

The stability of $Ni_3Al$ precipitates under irradiation was studied by NHM [15] in a Ni-Al alloy irradiated at 550°C with 100 keV Ni ions. It was noted that the stabilised state of precipitates, which depends on the balance between the irradiation dissolution of precipitates and their growth by irradiation enhanced diffusion, can be established to maintain the correct solute concentration in a dynamic solution.

In the mean-field approximation used in [15], assuming uniform spatial distribution of particles of radius $R$ and number density $N$, whose growth kinetics is described by Eq. (1), the requirement for conservation of total solute concentration takes the form,

$$c_A = \frac{4}{3}\pi R^3 N c_p + \bar{c}, \tag{68}$$

where $c_p$ is the concentration of solute in the precipitates, $c_A$ is the total concentration of solute, which after substituting into the condition of balance between the irradiation dissolution of precipitates and their growth derived from Eq. (3),

$$\bar{c} - c_s - \frac{\xi K_0}{D} R = 0, \tag{69}$$

leads to the relationship,

$$\frac{4}{3}\pi c_p R^3 N + R \frac{\xi K_0}{D\bar{c}} + (c_s - c_A) = 0, \tag{70}$$

whose solution determines the stabilised value $R_{st}$ at a fixed value of $N$. If, therefore, a system has a given density of coarse precipitates (with $R > R_{st}$), irradiation will cause them to shrink, whilst fine scale precipitates (with $R < R_{st}$) will grow (so-called 'inverse Ostwald ripening', cf. [1]).

In accordance with numerical analysis of NHM, $R_{st}$ decreases monotonically with increasing the fixed value of $N$. However, as noted in [15], in practice, it is unlikely that the precipitate number density will remain constant, since as the particles dissolve, the concentration of the solution will increase, leading to the nucleation of new particles from the solution. Eventually, all precipitates should attain the stabilised size defined by Eq. (70) at a number density set by irradiation and adjusted to maintain the correct concentration in the dynamic solution, which, however, cannot be determined using NHM theory.

This problem can be resolved by recognising that the decrease in $R_{st}$ (which at time $t$ is a function of the current value of $N(t)$) will terminate once the threshold radius $\hat{R}$ is attained along with the threshold supersaturation $\hat{S}$, below which the nucleation of new precipitates becomes unfeasible (as shown above in Section 3).

For simplicity, it will be assumed that $4\pi R_{st}^3 N_{st}/3 \ll 4\pi \lambda^3 N_{st}/3 \ll 1$; in this case, the growth of each particle can be considered independently from other particles (due to the first inequality) and described by the simplified growth equation, Eq. (1) (due to the second inequality, as explained in Section 3, giving $N_{st} \ll 2 \cdot 10^{24}$ m$^{-3}$). Otherwise, a more sophisticated cell model (e.g. [42]) can be used, in which each particle (of the same size) is in a matrix cell of radius $L$ with $4\pi L^3 N_{st}/3 = 1$, which will require further modification of the present analysis (which is beyond the scope of this work).

The decrease in $R$ under irradiation can occur without changing the particle number density $N_0$ (as assumed in NHM calculations), if the supersaturation of the solution, established as a result of system stabilisation, does not exceed the threshold value $\hat{S}$. Indeed, in this case, during the dissolution of particles under irradiation, the concentration of the solution increases monotonically, remaining below the threshold, which prevents the formation of new particles, so that the particle number density remains unchanged and their size decreases until it reaches the stabilised value found for $N_0$ from Eq. (70). In particular, if the total concentration $c_A = S_A c_{e,\infty}$ is below the threshold value $\hat{c}_A = \hat{S} c_{e,\infty}$ (at a given temperature), nucleation of new particles (at this temperature) becomes impossible.

However, in a more general case, the dissolution of particles can be accompanied with the nucleation of new particles, so that their number density $N(t)$ increases monotonically (in accordance with observations in [15]). This leads to a decrease in the stable particle radius $R_{st}(N(t))$ (calculated from Eq. (70) for each value $N(t)$) along with a variation in the stable solution concentration $\bar{c}_{st}(N(t))$ calculated from Eq. (69) as $\bar{c}_{st}(N(t)) = c_s + (\xi K_0/D) R_{st}(N(t))$ with $c_s = c_{e,\infty} \exp\left[3c_{e,\infty}(S-1) + \frac{2\gamma v}{kTc_p R}\right]$ (see Appendix B), until the threshold values of the supersaturation and radius (at which nucleation of new particles ceases) are reached, $S_{st} = \hat{S}$ and $R_{st} = \hat{R}$. This condition determines a complete stabilisation of the system, characterised by a balance between the dissolution and growth of existing particles in the absence of nucleation of new particles.

Correspondingly, the stabilised particle number density, $N_{st}$, is determined by the equation obtained by substituting $R = \hat{R}$ into Eq. (70),

$$\frac{4}{3}\pi c_p \hat{R}^3 N_{st} + \hat{R}\frac{\xi K_0}{D\bar{c}_{st}} + (c_s - c_A) = 0, \tag{71}$$

where $\hat{R}$ satisfies the zero growth rate condition derived from Eq. (B.9) (at $\dot{x} = 0$),

$$1 - S^{-1} \exp\left(3c_{e,\infty}(S-1)\right) \exp\left(\frac{2\gamma v}{kTc_p \hat{R}}\right) - \frac{\xi K_0}{DSc_{e,\infty}} \hat{R} = 0, \tag{72}$$

which, when substituted into Eq. (71), gives

$$N_{st} = \frac{3}{4\pi c_p \hat{R}^3}\left(S_A - \hat{S}\right), \tag{73}$$

where $S_A = c_A/c_{e,\infty}$, and $\hat{R}$ and $\hat{S}$ are calculated in Appendix B, Eqs (B.13) and (B.14).

For Ni$_3$Al precipitates in Ni-Al alloys studied in [15] at $T \approx 550°C$, $K_0 \approx 10^{-2}$ s$^{-1}$, and $D \approx 6 \cdot 10^{-18}$ m$^2$/s (corresponding to $\rho_d \sim 10^{15}$ m$^{-2}$), the threshold radius and supersaturation are estimated as $\hat{R} \approx 0.8$ nm and $\hat{S} \approx 1.3$ (see Section 3), and thus, using these and other parameters from Section 3, Eq. (73) shows that a stabilised value $N_{st}$ exists, if $c_A \geq \hat{c}_A = \hat{S}c_{e,\infty} \approx 0.13$. In particular, for $c_A \approx 0.135$ (specified in [15]), complete stabilisation occurs at $N_{st} \approx 5 \cdot 10^{25}$ m$^{-3}$, which, however, was not achieved during the NHM testing period.

As the temperature decreases, the lower limit $\hat{c}_A$ increases and may exceed $c_A = 0.135$, at which stabilisation becomes impossible. Consequently, there is a certain critical temperature (at a given $c_A$), below which the precipitates dissolve and above which they persist, in accordance with experimental observations [15]. This critical temperature can be evaluated (as $\approx 330°C$) from Fig. 1, calculated using Eqs (B.13), (B.14) and (73) at $\rho_d = 2 \cdot 10^{15}$ m$^{-2}$ and $\gamma = 0.01$ J/m$^2$.

Given that Ni$_3$Al precipitates nucleate in a chemically disordered fcc Ni matrix with very little lattice mismatch, allowing them to form a coherent interface, these results should be considered as qualitative estimates that can be improved by replacing $\gamma$ (in the expressions for $\hat{R}$ and $\hat{S}$) with the (poorly defined) effective surface energy and tuning it to better fit the experimental data (as discussed above in Section 5).

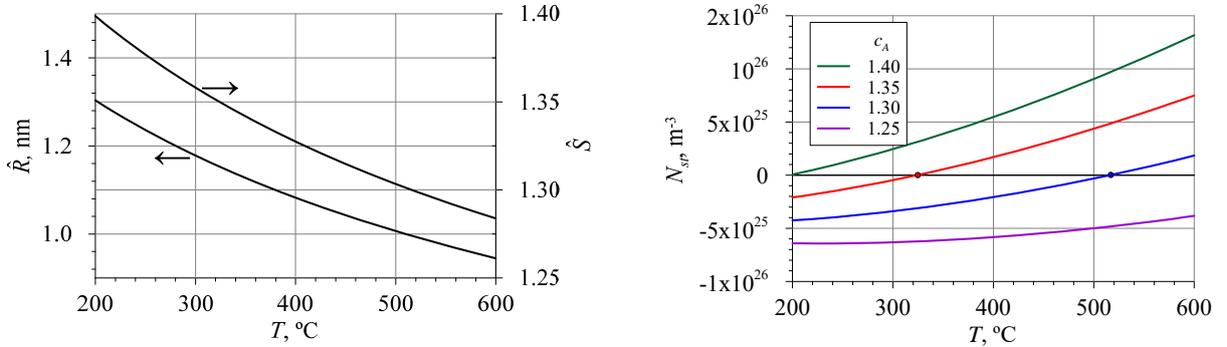

Fig. 1. Threshold radius $\hat{R}$ and supersaturation $\hat{S}$ (*left*), and stabilised particle number density $N_{st}$ (*right*), calculated using Eqs (B.13), (B.14) and (73) at $\rho_d = 2 \cdot 10^{15}$ m$^{-2}$ and $\gamma = 0.01$ J/m$^2$, roughly corresponding to the conditions of the NHM tests [15].

The kinetic description of NHM tests could be performed using the new model, which takes into account the nucleation and growth kinetics of precipitates described by Eqs (4) and (28) and the stabilisation of the system described by Eq. (73), once the model is implemented in the fuel performance code (e.g., BERKUT [43], as foreseen in the near future).

## 7. Discussion

In order to elucidate the experimental observations pertaining to the formation of precipitate phases in irradiated alloys [1–4], a number of thermodynamic models have demonstrated that the solubility limit of the solid solution can be elevated by increasing the (non-equilibrium) concentration of vacancies in the solid solution [5, 6]. This mechanism may act concomitantly with ballistic mixing of atoms in the solution, which was proposed in [7] as an additional mechanism for increasing of the solubility limit in the matrix under irradiation.

These predictions correlate with results of Russel's model for nucleation of incoherent precipitates [8], which showed that excess vacancies can shift the critical supersaturation and thus destabilise undersized precipitate phases or stabilise oversized ones. However, as was later shown in [12], the effect of vacancies is strongly attenuated by excess interstitials formed under irradiation conditions together with vacancies, and therefore this mechanism often becomes insufficient to explain the observed inhibition of new phase formation [14]. For this reason, it was suggested in [14] that irradiation prevents the formation of some phases by bombardment processes that physically knock the precipitate atoms into solution.

The effect of the ballistic re-solution of solute atoms from precipitates into the surrounding matrix was studied by Nelson, Hudson and Mazey (NHM) [15], who, based on their experimental observations, developed a model for the precipitate growth kinetics that accounts for dissolution by recoil and disordering processes. It was shown in [15] that if the system has a given density of coarse precipitates, irradiation will cause them to shrink and at the same time, existing fine scale precipitates will grow (so-called 'inverse Ostwald ripening'), until the precipitates stabilise at a certain size.

Since, in line with classical nucleation theory, the nucleation kinetics of precipitates is closely related to their growth mechanism, it can be expected that the ballistic re-solution will also affect the nucleation mechanism (as actually suggested in [14]). Accordingly, in the present work, a modification of classical nucleation theory was carried out as applied to solid solutions under irradiation, taking into account the influence of ballistic re-solution on the nucleation kinetics of precipitates.

Within the framework of classical nucleation theory, the nucleation of unary precipitates in a binary solid solution is represented by the translation of clusters ($x$-mers) in the phase space of cluster size $x$, with the nucleation rate calculated as the steady state flux $J_x$ of clusters between adjacent size classes in the phase space $x$. In this approach, the zero flux condition, $J_x = 0$, which represents the detailed balance between gaining and loss of monomers (solute atoms) by clusters, determines an equilibrium size distribution function, whose extremum defines the critical nucleus size $x^*$.

Under steady state irradiation, the zero flux condition determines a 'quasi-equilibrium' size distribution function (in steady state rather than equilibrium), which can be represented as $\rho_0(x) = C \exp[-\Delta \bar{G}_0(x)/kT]$, where $\Delta \bar{G}_0(x)$ is a non-thermodynamic function that is a 'quasi-equilibrium' analogue of the free energy of cluster formation in the absence of irradiation. This function can be calculated from the detailed balance condition (in discrete form) using the Fuchs algorithm [29], taking into account the ballistic term in the loss rate from clusters.

As a result, it is shown that $\Delta \bar{G}_0(x)$ can have two extrema (if the supersaturation ratio $S$ is high enough); the first is an unstable extremum that defines the critical cluster radius, and the second is a stable extremum that defines the maximum particle radius at which supercritical particle growth ceases. With a decrease in $S$, the two extrema approach each other until they merge together at some radius $\hat{R}$ when a threshold supersaturation $\hat{S}$ is reached, below which nucleation is unfeasible. In particular, this threshold effect may explain the inhibition of new phase formation observed in experiments, if the supersaturation ratio in the solid solution is below the threshold value, $1 \leq S < \hat{S}$.

The ballistic re-solution mechanism operates alongside the effects of excess point defects, considered previously in [12] for incoherent particles and in [38] for coherent particles, which have been incorporated into the new model to describe nucleation under irradiation in a more consistent manner. It is shown that for incoherent particles, the influence of excess point defects becomes significant at relatively low irradiation temperatures, whereas at high irradiation temperatures, ballistic effects make a key contribution to the suppression of the nucleation rate and the reduction of the stability range of the new phase.

For coherent particles, the elastic strain energy, which, as opposed to the case of incoherent particles, is not compensated by point defects, has an effect on the stability of the new phase, shifting it to higher supersaturations. Besides, the rate of coherent particle growth, which is controlled by the relatively slow nucleation of two-dimensional terraces on an atomically smooth particle interface, is strongly suppressed, which leads to a significant reduction in the nucleation rate [40]. Under irradiation conditions, this reduction in the nucleation rate is amplified by the ballistic resolution effect, but can be efficiently compensated by a decrease in the effective interface surface energy $\gamma_{eff}$ caused by adsorption and mutual annihilation of excess point defects at the particle interface, which can significantly reduce the nucleation barrier $\Delta G_0(x^*)$ [38]. Ultimately, these effects can lead to considerable uncertainty in the value of the nucleation barrier, which for practical applications can be determined by tuning it (e.g. by changing $\gamma_{eff}$) to better match the experimental data.

The developed model was modified to consider stoichiometric binary precipitates in a binary solid solution and applied for qualitative interpretation of results of NHM tests [15], which studied the stability of Ni$_3$Al precipitates in Ni-Al alloys under irradiation. In accordance with the NHM model developed on the base of these tests, all precipitates should attain the stabilised size at a number density set by irradiation and adjusted to maintain the correct concentration in the dynamic solution, which, however, cannot be determined using the NHM model.

This problem can be solved using the new model showing that the decrease in particle radius and increase in their number density cease once the threshold radius $\hat{R}$ and supersaturation $\hat{S}$ are reached. It was shown that the nucleation of precipitates is not possible if the supersaturation ratio $S$ of the alloy is below the threshold value, $1 \leq S < \hat{S}$, whereas at higher supersaturations, fine scale precipitates form and grow (along with the shrinkage of existing coarse precipitates, until the system stabilises completely). This allows estimating the threshold concentration for an alloy, $\hat{c}_A = \hat{S}c_{e,\infty}$, below which the stabilisation of precipitates becomes not possible and they dissolve completely. Since the supersaturation threshold depends on temperature $T$, different regimes (with $S > \hat{S}(T)$ and $S < \hat{S}(T)$) can be realised in the same alloy at different temperatures, in qualitative agreement with observations [15]. A quantitative kinetic description of the tests could be performed using the new model, once it is implemented in the fuel performance code.

## 8. Conclusions

The NHM model [15] for the growth kinetics of precipitates under irradiation conditions, which accounts for dissolution of solute atoms from precipitates into the surrounding matrix by ballistic recoil and disordering processes, was analysed and used to develop a nucleation model. Given that, according to classical nucleation theory, the nucleation kinetics of precipitates is closely related to their growth mechanism, it can be expected that the ballistic re-solution would also influence the nucleation mechanism (as also suggested in [14]). Accordingly, in the present work, a modification of classical nucleation theory was carried out as applied to the nucleation of unary precipitates in solid solutions under irradiation, taking into account the influence of ballistic re-solution on the nucleation kinetics. It was shown that there is a threshold value of supersaturation $\hat{S}$, depending on the irradiation intensity and temperature, below which nucleation of precipitates is unfeasible. In particular, this can explain the observed suppression of new phase formation under irradiation in the case of $1 \leq S < \hat{S}$.

The ballistic re-solution mechanism operates alongside the effects of excess point defects formed under steady irradiation conditions, considered previously for incoherent and coherent particles, which have been incorporated into the new model to describe nucleation in a more consistent manner.

As a result, it is demonstrated that, for incoherent particles, ballistic effects make the main contribution to the suppression of the nucleation rate and stability of the new phase under irradiation at relatively high temperatures. For coherent particles, ballistic re-solution amplifies the strong decrease in nucleation rate caused by a slow mechanism of the particle growth (by two-dimensional nucleation of terraces on an atomically smooth particle interface), which can be effectively compensated by a reduction in the nucleation barrier caused by adsorption and mutual annihilation of point defects at the particle interface.

The developed model was modified to consider stoichiometric binary precipitates and applied to qualitatively interpret the results of NHM tests [15], in which the stability of $\gamma'$ ($Ni_3Al$) precipitates in Ni-Al alloys under irradiation was studied. It was shown that the decrease in particle radius and increase in their number density observed during the tests cease once the threshold supersaturation $\hat{S}$ and radius $\hat{R}$ are reached. This allows estimating the maximum concentration value for the alloy under irradiation, above which the complete stabilisation of precipitates can be reached, and below which they dissolve. Since the supersaturation threshold depends on temperature, different regimes (with $S > \hat{S}(T)$ and $S < \hat{S}(T)$) can be realised in the same alloy at different temperatures, in qualitative agreement with observations [15].

**Acknowledgements**

Dr. V. Tarasov (IBRAE, Moscow) is acknowledged for valuable discussions and assistance in numerical calculations (Fig. 1).

**Appendix A**

Eq. (10) can be modified using the series expansion of $\beta(x-1)$,

$$\beta(x-1)s_{x-1} \approx \beta(x)s_x - \frac{d(\beta(x)s_x)}{dx} + \frac{1}{2}\frac{d^2(\beta(x)s_x)}{dx^2} - \frac{1}{6}\frac{d^3(\beta(x)s_x)}{dx^3}\ldots, \tag{A.1}$$

where $\beta(x)s_x \propto x^{1/3}$. Therefore, for clusters with $x \gg 1$ (to which classical nucleation theory applies), this leads to $\frac{d(\beta(x)s_x)}{dx} = \frac{1}{3x}\beta(x)s_x \ll \beta(x)s_x$, and $\frac{d^2(\beta(x)s_x)}{dx^2} = \frac{2}{9x^2}\beta(x)s_x \ll \frac{d(\beta(x)s_x)}{dx} \ll \beta(x)s_x$, etc., and thus with sufficient accuracy, Eq. (A.1) reduces to

$$\beta(x-1)s_{x-1} \approx \beta(x)s_x. \tag{A.2}$$

In this approximation, Eq. (10) takes the form,

$$\beta(x)s_x\rho_0(x-1) - [\alpha^{(th)}(x) + \alpha^{(irr)}(x)]s_x\rho_0(x) \approx 0, \tag{A.3}$$

which justifies Eq. (20).

However, as Eq. (20) applies only to 'macroscopic' clusters with $x \gg 1$, a more appropriate expression, instead of Eq. (21), is

$$\frac{\rho_0(x_0)}{\rho_0(x^*)} = \frac{\rho_0(x_0)}{\rho_0(x_0+1)}\frac{\rho_0(x_0+1)}{\rho_0(x_0+2)}\ldots\frac{\rho_0(x^*-1)}{\rho_0(x^*)} \approx \prod_{i=x_0}^{x^*} S^{-1}\left[\exp\left(\frac{2\gamma v}{R_i kT}\right) + \frac{\xi K_0 R_i}{Dc_{e,\infty}}\right], \tag{A.4}$$

where $1 \ll x_0 \ll x^*$. Due to the second inequality, $x_0 \ll x^*$, one can neglect in Eq. (A.4) the ballistic re-solution rate compared to the thermal loss rate for a particle of size $x_0$ in a wide range of supersaturations $S \geq 1$. Indeed, for the critical radius $R^*$, which can be estimated from Eq. (31) as

$$R^* \equiv R_1^* = \frac{Dc_{e,\infty}(S-1)}{2\xi K_0}\left\{1 - \left[1 - \frac{8\gamma v\xi K_0}{kTDc_{e,\infty}(S-1)^2}\right]^{1/2}\right\} \leq \frac{Dc_{e,\infty}(S-1)}{2\xi K_0}, \tag{A.5}$$

one can estimate

$$\frac{\alpha^{(th)}(x^*)}{\beta(x^*)} = \exp\left(\frac{2\gamma v}{kTR^*}\right) \geq \exp\left(\frac{4\gamma v\xi K_0}{kTc_{e,\infty}(S-1)}\right) \geq 1, \tag{A.6}$$

whereas

$$\frac{\alpha^{(irr)}(x^*)}{\beta(x^*)} = \frac{R^*\xi K_0}{Dc_{e,\infty}} \leq \frac{(S-1)}{2}, \tag{A.7}$$

and hence, the inequality $\alpha^{(th)}(x^*) \geq \alpha^{(irr)}(x^*)$ holds for $S \leq 3$. Consequently, for $x_0 \ll x^*$, the inequality $\alpha^{(th)}(x_0) \gg \alpha^{(irr)}(x_0)$ holds (at least) in the range $1 \leq S \leq 3$, given that $\alpha^{(th)}(x_0)/\beta(x_0) \gg \alpha^{(th)}(x^*)/\beta(x^*)$ and $\alpha^{(irr)}(x_0)/\beta(x_0) \ll \alpha^{(irr)}(x^*)/\beta(x^*)$.

This means that in this range, $\rho_0(x_0)$ can be approximated with good accuracy by the equilibrium size distribution function (defined in the absence of ballistic re-solution),

$$\rho_0(x_0) \approx N_1 \exp[-\Delta G_0(x_0)/kT] = N_1 \exp\left(x_0 \ln S - \frac{3}{2}ax_0^{2/3}\right), \tag{A.8}$$

where $\Delta G_0(x)$ is defined in Eq. (7), and the pre-exponential factor is $N_1$, according to the Frenkel model [48] (critically discussed in the literature, see, e.g. [49], but additionally justified in the author's paper [33] and more accurately in Appendix C to [38]).

Therefore, similar to the derivation of Eq. (24) from Eq. (21), it can be deduced from Eq. (A.4),

$$\ln\frac{\rho_0(x_0)}{\rho_0(x^*)} \approx \int_{x_0}^{x^*}\left\{\ln\left[S^{-1}\exp\left(\frac{2\gamma v}{kTR(y)}\right)\right] + \ln\left[1 + \frac{\xi K_0}{Dc_{e,\infty}}\left(\frac{3v}{4\pi}\right)^{1/3} y^{\frac{1}{3}}\exp\left(-ay^{-\frac{1}{3}}\right)\right]\right\} dy, \tag{A.9}$$

or

$$\rho_0(x^*) \approx \rho_0(x_0)\exp\left[(x^* - x_0)\ln S - \frac{3}{2}a\left(x^{*\frac{2}{3}} - x_0^{\frac{2}{3}}\right) - I(x^*, x_0)\right], \tag{A.10}$$

where

$$I(x, x_0) = \int_{x_0}^{x}\ln\left[1 + \frac{\xi K_0}{Dc_{e,\infty}}\left(\frac{3v}{4\pi}\right)^{1/3} y^{\frac{1}{3}}\exp\left(-ay^{-\frac{1}{3}}\right)\right] dy, \tag{A.11}$$

which after substitution of Eq. (A.8) into Eq. (A.10) takes the form

$$\rho_0(x^*) \approx N_1 \exp\left(x^* \ln S - \frac{3}{2}ax^{*\frac{2}{3}} - I(x^*, x_0)\right). \tag{A.12}$$

Taking into account that $I(x, x_0) = I(x, 0) - I(x_0, 0) = I(x) - I(x_0)$, where $I(x) \equiv I(x, 0)$ is defined in Eq. (26), and that $I(x_0) \ll I(x^*)$ at $x_0 \ll x^*$, as seen from Eqs (37) and (38) (with $c_p = 1$), the lower limit in the integral of Eq. (A.11), can be replaced with good accuracy by 0 leading to

$$\rho_0(x^*) \approx N_1 \exp\left(x^* \ln S - \frac{3}{2}ax^{*\frac{2}{3}} - I(x^*)\right), \tag{A.13}$$

which justifies Eq. (27) at least in the range $1 \leq S \leq 3$, which may cover a significant part of the metastable zone of the equilibrium phase diagram (outside the spinodal instability zone).

**Appendix B**

Modification of the nucleation theory to the case of stoichiometric binary Ni$_3$Al precipitates in Ni-Al substitutional alloy can be performed as follows. The structure of Ni$_3$Al intermetallic compound is characterized by a face centred cubic (fcc) arrangement (where Al atoms occupy the corners of the cube, and Ni atoms occupies the face centres). Correspondingly, the Gibbs free energy of formation of a nucleus consisting of $x$ Al atoms and $3x$ Ni atoms takes the form,

$$\Delta G_0(x) = x\left(\mu_{Al}^{(p)} + 3\mu_{Ni}^{(p)} - \mu_{Al}^{(m)} - 3\mu_{Ni}^{(m)}\right) + 4\pi R^2 \gamma, \tag{B.1}$$

where $\mu_i^{(p)}$ and $\mu_i^{(m)}$ are chemical potentials of components $i = $ Al, Ni in the precipitate phase and in the matrix, respectively; $R = (3vx/4\pi c_p)^{1/3}$ is the radius of the precipitate (with the concentration of Al atoms $c_p = 0.25$), $v$ is the mean volume occupied by atoms of the two types in the compound.

Given a low concentration of Al atoms in the Ni matrix, $c \equiv c_{Al} \ll 1$, the weak solution approximation for this solid solution can be properly used, giving $\mu_{Al}^{(m)} = \mu_{Al,0}^{(m)} + kT \ln c$ and $\mu_{Ni}^{(m)} = \mu_{Ni,0}^{(m)} - kTc$, where $\mu_{Al,0}^{(m)}$ and $\mu_{Ni,0}^{(m)}$ are the standard chemical potentials of solute (Al) and solvent (Ni) atoms, respectively (cf., e.g. [28]). Therefore, the equilibrium at the flat interface between the solid solution and Ni$_3$Al phase is described by the relations $\mu_i^{(p)} = \mu_i^{(m)}$, or $\mu_{Al}^{(p)} = \mu_{Al,0}^{(m)} + kT \ln c_{e,\infty}$ and $\mu_{Ni}^{(p)} = \mu_{Ni,0}^{(m)} - kTc_{e,\infty}$, where $c_{e,\infty}(T)$ is the equilibrium concentration (from the Ni-Al phase diagram) defined in Section 2, which determines the supersaturation ratio as $S = c/c_{e,\infty}$. Substituting these relations into Eq. (B.1) leads to

$$\Delta G_0(x) = xkT[-\ln S + 3c_{e,\infty}(S-1)] + 4\pi\gamma \left(\frac{3v}{4\pi c_p}\right)^{2/3} x^{2/3}, \tag{B.2}$$

which in the case $S - 1 \ll 1$ (well corresponding to the conditions of the tests [15]) can be simplified (for convenience) as

$$\Delta G_0(x) = -xkT(1 - 3c_{e,\infty})\ln S + 4\pi\gamma \left(\frac{3v}{4\pi c_p}\right)^{2/3} x^{2/3}. \tag{B.3}$$

which differs from Eq. (7) only by the coefficients of its two terms.

Note that Eq. (B.1) can be equivalently represented as

$$\Delta G_0(x) = x\left(\mu_{Ni_3Al}^{(p)} - \mu_{Ni_3Al}^{(m)}\right) + 4\pi R^2 \gamma, \tag{B.4}$$

where $\mu_{Ni_3Al}^{(p)}$ and $\mu_{Ni_3Al}^{(m)}$ are chemical potentials of Ni$_3$Al 'monomers' in the precipitate phase and in the matrix, respectively, which obey the equilibrium condition in each phase, $3Ni + Al \leftrightarrow Ni_3Al$, or $\mu_{Ni_3Al}^{(p,m)} = \mu_{Al}^{(p,m)} + 3\mu_{Ni}^{(p,m)}$. This facilitates the application of classical nucleation theory, in which nucleation of the Ni$_3$Al phase is represented by the translation of clusters (consisting of Ni$_3$Al monomers) in phase space of cluster size $x$ (see Section 3), and the nucleation rate is defined as a steady state flux $J_x$ of such clusters between adjacent size classes in phase space.

Accordingly, the pre-exponential factor in the equilibrium size distribution function included in Eq. (17), is derived by considering Ni$_3$Al clusters (with the number density $N_x$), formed by Ni$_3$Al monomers (with the number density $N_1$) in the Ni matrix (with the number density of lattice sites $N_0 \approx v^{-1}$),

$$\rho_0(x) \approx N_1 \exp[-\Delta G_0(x)/kT], \tag{B.5}$$

(see Appendix C to [38] for details of the derivation for such systems), whereas under irradiation conditions it is modified to the form,

$$\rho_0(x) \approx N_1 \exp[-\Delta \bar{G}_0(x)/kT], \tag{B.6}$$

where $\Delta \bar{G}_0(x)$ additionally takes into account ballistic re-solution from the cluster (cf. Section 3).

The number density of Ni$_3$Al monomers in the matrix, $N_1$, is calculated from the equilibrium condition in the matrix, $\mu_{Ni_3Al}^{(m)} = \mu_{Al}^{(m)} + 3\mu_{Ni}^{(m)}$, where $\mu_{Ni_3Al}^{(m)} \approx \mu_{Ni_3Al,0}^{(m)} + kT \ln(N_1/N_0)$, and $\mu_{Al}^{(m)} \approx \mu_{Al,0}^{(m)} + kT \ln(N_{Al}/N_0) \approx \mu_{Al,0}^{(m)} + kT \ln c$, and $\mu_{Ni}^{(m)} \approx \mu_{Ni,0}^{(m)} - kTc \approx \mu_{Ni,0}^{(m)}$, resulting in

$$N_1 = K_d^{-1} c N_0 = K_d^{-1} S c_{e,\infty} v^{-1}, \tag{B.7}$$

where $K_d = \exp\left[\left(\mu_{Ni_3Al,0}^{(m)} - \mu_{Al,0}^{(m)} - 3\mu_{Ni,0}^{(m)}\right)/kT\right]$ is the dissociation constant (which can be estimated by following the procedure described in Appendix C to [50]).

The kinetic factor in the nucleation rate, Eq. (17), is determined by the diffusion transport of Al atoms to the critical cluster and is calculated (by following the procedure described in Appendix B to [38]) as

$$\omega^* = 4\pi R^{*2} \beta(x^*) = \left(\frac{4\pi}{v}\right)^{2/3} \left(\frac{3}{c_p}\right)^{1/3} D_{Al} S c_{e,\infty} x^{*1/3}, \tag{B.8}$$

where $\beta(x) = \frac{D_{Al} c_{Al}}{vR} = \frac{D_{Al} c_{Al}}{v}\left(\frac{4\pi c_p}{3vx}\right)^{1/3} = D_{Al} S c_{e,\infty}\left(\frac{4\pi c_p}{3v^4 x}\right)^{1/3}$, $c_p = 0.25$, and $R = \left(\frac{3vx}{4\pi c_p}\right)^{1/3}$.

Taking into account these corrections, the resulting equations in Section 3 are modified as follows,

$$\dot{x} = 4\pi R^2 \beta(x) \left[1 - S^{-1} \exp\left(3c_{e,\infty}(S-1) + \frac{2\gamma v}{kT c_p R}\right) - \frac{\xi K_0}{DSc_{e,\infty}} R\right], \tag{B.9}$$

$$Z = \left[-\frac{1}{2\pi kT} \frac{d^2 \Delta \bar{G}_0(x^*)}{dx^2}\right]^{\frac{1}{2}} =$$

$$\left\{\frac{1}{6\pi}\left[-a' c_p^{-\frac{1}{3}} x^{*-\frac{4}{3}} + S'^{-1} \frac{\xi K_0}{Dc_{e,\infty}}\left(\frac{3v}{4\pi c_p}\right)^{\frac{1}{3}} \left(1 + a' c_p^{\frac{2}{3}} x^{*-\frac{1}{3}}\right) x^{*-\frac{2}{3}} \exp\left(a' c_p^{-\frac{1}{3}}(1-c_p) x^{*-\frac{1}{3}}\right)\right]\right\}^{\frac{1}{2}}. \tag{B.10}$$

$$\dot{N} \approx K_d^{-1} (4\pi)^{\frac{2}{3}} \left(\frac{3}{c_p}\right)^{\frac{1}{3}} D_{Al} (Sc_{e,\infty})^2 v^{-5/3} Z \exp\left[x^* \ln S' - \frac{3}{2} \frac{a'}{c_p^{1/3}} x^{*\frac{2}{3}} - I'(x^*)\right], \tag{B.11}$$

$$I'(x) = \int_0^x \ln\left[1 + \frac{\xi K_0}{Dc_{e,\infty}}\left(\frac{3v}{4\pi c_p}\right)^{1/3} y^{\frac{1}{3}} \exp\left(-a' c_p^{2/3} y^{-\frac{1}{3}}\right)\right] dy, \tag{B.12}$$

where $S' = S^{(1-3c_{e,\infty})}$, and $a' = \left(\frac{4\pi}{3c_p}\right)^{1/3} \frac{2\gamma v^{2/3}}{kT}$, and equations for the threshold (under the condition $\frac{2\gamma v}{kT c_p R} \ll 1$) take the form

$$\hat{S}' = \hat{S}^{(1-3c_{e,\infty})} \approx 1 + \left(\frac{8\gamma v \xi K_0}{kTDc_p c_{e,\infty}}\right)^{\frac{1}{2}}, \tag{B.13}$$

$$\hat{R} \approx \left(\frac{2\gamma v D c_{e,\infty}}{kT\xi K_0 c_p}\right)^{1/2}. \tag{B.14}$$